\documentclass[aps,showpacs,amsmath,amssymb]{revtex4}

\usepackage{graphicx}
\usepackage{epsfig}

\begin{document}
 
\title{On ``the authentic damping mechanism'' of the phonon damping model}
 
\author{V.~Yu.~Ponomarev\cite{byline1}}
\affiliation{Institut f\"ur Kernphysik, Technische Universit\"at Darmstadt, 
D--64289  Darmstadt, Germany}
\date{\today }
 
\begin{abstract}
Some general features of the phonon damping model are presented.
It is concluded that the fits
performed within this model have no physical content.
\end{abstract}
 
 
\maketitle
 
In a recent article~\cite{d1}, the phonon damping model (PDM) has been 
applied for a description of the giant dipole resonance (GDR) and pygmy 
dipole resonance (PDR) in oxygen and calcium chains from double-magic
to exotic isotopes. It has been argued that it provides much better
agreement with the GDR photoabsorption cross sections (PCS) than more 
advanced, microscopic, approaches.
The main purpose of the present Comment is to understand why it is so.
 
The PDM is a model in which the mode under discussion, the phonon $Q$
(with the excitation energy $\omega$) and 
its coupling to $N$ uncorrelated $1p1h$ states are described 
phenomenologically. The $1p1h$ spectrum is calculated microscopically. 
Let us start with the PDM application to double-magic nuclei. 
 
A key starting point of almost all PDM calculations is an approximation that 
the phonon and any $1p1h$ state interact with an equal strength $f_1$, a model
parameter. From microscopic point of view, this assumption is very far from
reality.

The general features of the $Q$ fragmentation due its interaction 
with some other states $| \alpha \rangle$
may be found in textbooks (see, e.g., Appendix 2D in \cite{Boh69}).
Then, the second moment for the phonon distribution in the PDM has a 
simple analytical form:
\begin{equation}
W_2 = \left( f_1 \right)^2 \cdot N~.
\label{sm}
\end{equation}
Equation~(\ref{sm})
is exact and independent of the details of the spectrum $E_{\alpha}$.
However, the shape of the distribution does depend on it, having 
the Breit-Wigner (BW) form if the energies $E_{\alpha}$ are 
equidistant \cite{Boh69}. 
Again, the nature of $| \alpha \rangle$ 
(whether they are $1p1h$ or n$p$n$h$ states) is not 
essential. It is only important that the energy scales of 
$\omega$, $f_1$, and $E_{\alpha}$ are of the same order.
\begin{figure}[b]
\epsfig{figure=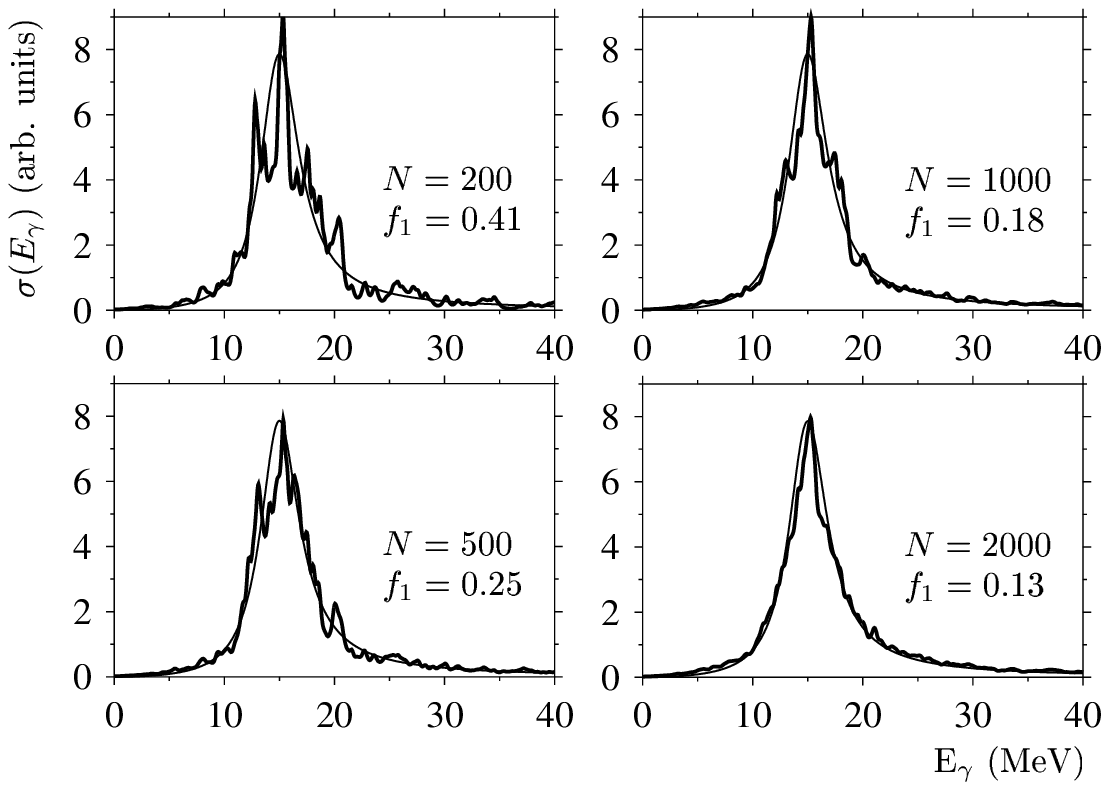,width=90mm,angle=0}
\caption{\label{fig1}
The PDM calculations of the GDR PCS with a random spectrum
$E_{\alpha}$ (thick line) in comparison with
the Lorentz distribution (thin line). 
See text for details.}
\end{figure}
  
This means that the BW form for the GDR within the PDM is a direct 
consequence of the assumption that the coupling matrix
element is the same for all $1p1h$ states. When a realistic $1p1h$ 
spectrum is used in the PDM calculations, the BW shape is disturbed. 
To check how strong is this disturbance in general, the PDM calculations 
with random values of $E_{\alpha}$ from 0 to 50~MeV have been performed.
The purpose of these calculations is to reproduce the Lorentz line for 
the GDR PCS in some hypothetical nucleus with $E_0 = 15.0$~MeV and $\Gamma =
4.3$~MeV by fitting the 
PDM parameters $f_1$ and $\omega$. 
The results of calculations with an additional smearing parameter
$\varepsilon = 0.5$~MeV (as in Ref.~\cite{d1} for the oxygen 
chain) are represented in Fig.~\ref{fig1} by thick lines. 
Cross sections are plotted in arbitrary units. For the amplitude adjustment, 
a free parameter $c_1$ is available in the PDM.
 
The calculations show that the PDM results for the GDR PCS converge rather
fast to the Lorentz line as $N$ increases, even for 
a random $E_{\alpha}$ spectrum. 
To exclude any accidental coincidence, the
calculations have been repeated with several different random spectra.
Qualitatively, the results are similar. So, any traces of the PDM 
``microscopy'' vanish if $N$ is not small.
 
Adopting the Lorentz shape for the GDR PCS as the model input, it is
not surprising that the PDM ``describes'' the photoabsorption
data better than microscopic models in which 
such physical observable as the GDR width is calculated(!).
But, to understand whether there is any physical content
behind the PDM fits, one needs to analyze the physical meaning of the
PDM parameters and/or check how it describes some independent data. 
  
In microscopic perturbative approaches, the matrix element of the 
interaction between  $1p1h$ configurations and a phonon tends to increase 
when a larger 
basis of $1p1h$ states is employed. This is due to the increase of the 
phonon's collectivity. However, in the PDM, the collectivity of
$Q$ does not depend  
on the $1p1h$ basis, and the strength parameter $f_1$ 
decreases with increasing N. Roughly, it 
goes as $f_1 \sim 1/\sqrt{N}$, since $W_2$
in Eq.~(\ref{sm}) is more or less fixed by the data to which $f_1$ is
adjusted. Since $f_1$ is determined not according to its physical
meaning, but only to fit the data, this procedure leads Dang {\it et al.}
into a contradiction in principle with rather general arguments on the 
properties
of the system under consideration. Indeed, the strength of the interaction
between any configuration $| \alpha_0 \rangle$ and $Q$ is 
determined not by their
physical properties but only by the number(!) of other configuration 
$| \alpha \rangle$.
The authors define it as ``microscopic description of damping''?!
 
Another misleading statement in \cite{d1} is that the coupling to 
higher-order graphs are included effectively in the strength parameter 
$f_1$ of the lowest order graphs. This is not true
because these are two different physical processes.
 
Let us briefly consider the PDM results in Ref.~\cite{d1}.
The physics of the essential difference of the GDR width in $^{40}$Ca and 
$^{48}$Ca is still an open question. Dang {\it et al.} report an
agreement with the data in both nuclei. The agreement for
$^{48}$Ca is obtained by renormalizing 
$f_1$ by 34\% (note, that $f_1$ is fitted up to 4 digits). 
But it is difficult to learn anything from this agreement when the
true physical meaning of $f_1$ in the PDM model is simply ignored.

The extention of the PDM to open-shell nuclei in \cite{d1} by including
pairing for the $1p1h$ states only stresses the internal PDM
problems. The lack of the PCS data for these nuclei, except for $^{18}$O,
allows Dang {\it et al.} to keep $f_1$ fixed from $^{18}$O to $^{24}$O, 
from $^{42}$Ca to $^{46}$Ca, and from $^{50}$Ca to $^{60}$Ca as an
assumption.
But the data available for $^{16}$O and $^{18}$O already
forces the authors to reduce $f_1$ by $\sim25\%$ from $^{18}$O
to $^{16}$O to achieve an agreement in both (see, \cite{note1}).
They claim that the renormalization is to compensate for the 
enlargement of the configuration space in $^{18}$O due to the pairing.  
But a smaller configuration space 
should lead to a larger(!) $f_1$ and not vice versa (see, Eq.~(\ref{sm})).
Again, considering the physical meaning of $f_1$, there are no physical
grounds for such renormalization.
 
The properties of the PDR are considered as independent data for the PDM
calculations. Although Dang {\it et al.} conclude a ``consistent and
quantitative description'' of this resonance, it is difficult to find any 
agreement of the calculation with the fine structure of the PCS at low 
energies presented in Fig.~4 of \cite{d1}, especially for $^{18}$O. 
For $^{40}$Ca and $^{48}$Ca, the high-resolution data below 10~MeV 
are available \cite{Har00}. The PDM results are compared to these
data in \cite{d1} for $^{48}$Ca, but not for $^{40}$Ca.
Such selective comparison may mislead the reader.
The PDM predictions for $^{40}$Ca were published before the data in 
Ref.~\cite{d3}. We find that the PDR exhausts 0.3\% of the EWSR in this 
nucleus.
The same value obtained experimentally equals in $^{40}$Ca 
about 0.007\% after the
two-phonon candidate  $[2^+_1 \times 3^-_1]_{1^-}$ at 6.950~MeV, which is 
outside the PDM space, is excluded from consideration.
The difference by a factor of 40 cannot be defined as quantitative
agreement.
 
The failure to describe the PDR by the GDR spreading to lower energies,
as the PDM does, 
has been sufficiently discussed in the literature (see, e.g.,
\cite{Har00,End00} as latest references). In microscopic models,
the PDR is associated with the excitation of the lowest
$1p1h$ $1^-$ configurations \cite{End00,Cat97,Oro98,Vre01,Col01}. 
These configurations are included in the PDM
model space but their B($E1$) values are set to zero to avoid
an obvious PDM problem with double counting.

To conclude, it is not clear what Dang {\it et al.} mean by the
``consistent and quantitative'' description of the GDR within the PDM
in \cite{d1}. A possibility to fit the PCS by  
the Breit-Wigner shape, which is
the model phenomenologic {\it ad hoc} input, is not under question.
For those nuclei for which the data is available and presented, 
the PDM needs different sets of the model parameters which are fitted to the 
described physical observables (three parameters for three observables). 
Taken together with the above analysis of the physical meaning of 
the strength parameter, this makes the physical content of the
PDM calculations very doubtful. The predictive power of this model is also
doubtful and there is no sense to use it for such purpose.
The nature of the PDR in the PDM contradicts the
microscopic understanding of this resonance and the conclusion that this
model describes the PDR properties on the quantitative level is 
not justified.
 
It is not possible to agree that the PDM fits confirm ``the authentic 
damping mechanism of giant resonances'' as ``the result of coupling between
collective phonon and non-collective $p$-$h$ configurations'' (with equal
matrix element).
 
The author thanks Professor J. Weil for a careful reading of the manuscript.


\begin{thebibliography}{99}
\bibitem{note1}
In Ref.~\cite{d1}, we read ``The values of PDM parameters for $^{16}$O are then
fixed in calculations for all oxygen isotopes'' (page 4).
\end{thebibliography}
\end{document}